\documentstyle[11pt,twoside,epsfig]{article}    

\newcommand{\beq}{\begin{equation}}      
\newcommand{\eeq}{\end{equation}}

\newlength{\dinwidth}  
\newlength{\dinmargin}          
\begin{document}      

\begin{titlepage}
\vspace*{1cm}
\begin{center}
\begin{LARGE}
\bf{ Structure Function  Measurements 
at HERA \\
\vspace*{0.3cm}
  and Perturbative QCD}
\end{LARGE}  
\vspace*{0.7cm}        

\begin{large}          
{Gregorio Bernardi}\\          
        on behalf of H1 and ZEUS collaborations
\footnote{Invited plenary talk given at Baryons '95
held in Sante F\'e, New Mexico, in October 1995.
The results presented here have been updated according to
the recent publications of  ZEUS~\cite{ZEUSF294},
 H1~\cite{H1F294,H1CC94} and E665~\cite{E665}.
Original figures can be found for instance in~\cite{Greg}.}
\\    
 LPNHE-Paris, 4 place Jussieu, 75252 Paris Cedex 05, France     
\end{large}
\end{center}

          

\vspace*{0.3cm}

\begin{abstract}  
\noindent
New results from the H1 and ZEUS collaborations
on the   measurement of cross-sections at very high $Q^2$ (up to 25000 GeV$^2$)
and
on the proton structure function $F_2(x,Q^2)$ 
for momentum transfers squared $Q^2~\geq$ 1.5~GeV$^2$  
 and  Bjorken $x \geq$ 3.5$\cdot 10^{-5}$ are reported,   
using data collected at HERA  mainly in 1994.       
No deviations from the Standard Model have been observed at high $Q^2$ and
$F_2$ is seen to increase significantly with decreasing $x$,   
even in the lowest reachable $Q^2$ region.  Comparisons at low $Q^2$ with
fixed target experiments and with  models based on  pomeron 
exchange are presented.  
The $F_2$ results are well described by a Next to Leading Order QCD fit,   
and are consistent at the  
 present level of precision with the rise at low $x$       
within this $Q^2$ range generated  via the DGLAP        
evolution equations. The gluon density is extracted and being observed
to rise at low $x$.  
\end{abstract}    

          
\section{Introduction}    
The HERA $ep$ collider has been designed to study Deep Inelastic Scattering
(DIS) at very high $Q^2$, where the strength of the electromagnetic and weak
forces become comparable and  
 where substructure of quarks might be observed.
However in the first 3 years of data operation, which allowed a steady growth
towards the design luminosity of the machine,
 most of the interest has focused on the study
of low $x$, low $Q^2$ DIS, where new tests of perturbative QCD can be 
performed. 
The first observations on the 1992 data showed a rise of the proton      
structure function $F_2(x,Q^2)$ at low $x < 10^{-2}$ with decreasing    
$x$~\cite{H1F292,ZEUSF292}, which was     
confirmed with the more precise data of 1993     
\cite{H1F293,ZEUSF293}.   
Such a behaviour is qualitatively expected in the double leading log    
limit of Quantum Chromodynamics~\cite{ALVARO}. It is, however, not      
clarified 
whether the linear QCD evolution equations, as the    
conventional DGLAP evolution~\cite{DGLAP} in $\ln Q^2$ and/or the BFKL  
evolution~\cite{BFKL} in $\ln(1/x)$, describe the rise of $F_2$ or      
whether there is a significant effect due to 
      non-linear parton recombination \cite{GLR}.
At low $Q^2$ ($\leq$ 5 GeV$^2$) the new results can be confronted 
with Regge inspired models, which expect a rather flat behaviour as a function
of $x$, in order to study the transition between  
\end{titlepage}          
\noindent
DIS and
 photoproduction.
The 1994 data have made possible to reach an extended
kinematic region, both in $Q^2$ and in $x$, and to confirm the 
persistance of the rise at low $x$ at the lowest 
$Q^2$ measured. The very high $Q^2$ results are 
presented in section 1.   
The latest $F_2$ measurements 
which have been achieved by using dedicated data samples are presented 
in section 2 and  their low $Q^2$ behaviour is discussed in section
 3. They are 
 analyzed in terms of perturbative QCD in section 4.

\section{Differential Cross-Sections at Very High $Q^2$}

\begin{figure}[htbp]       
\begin{center}    
\epsfig{file=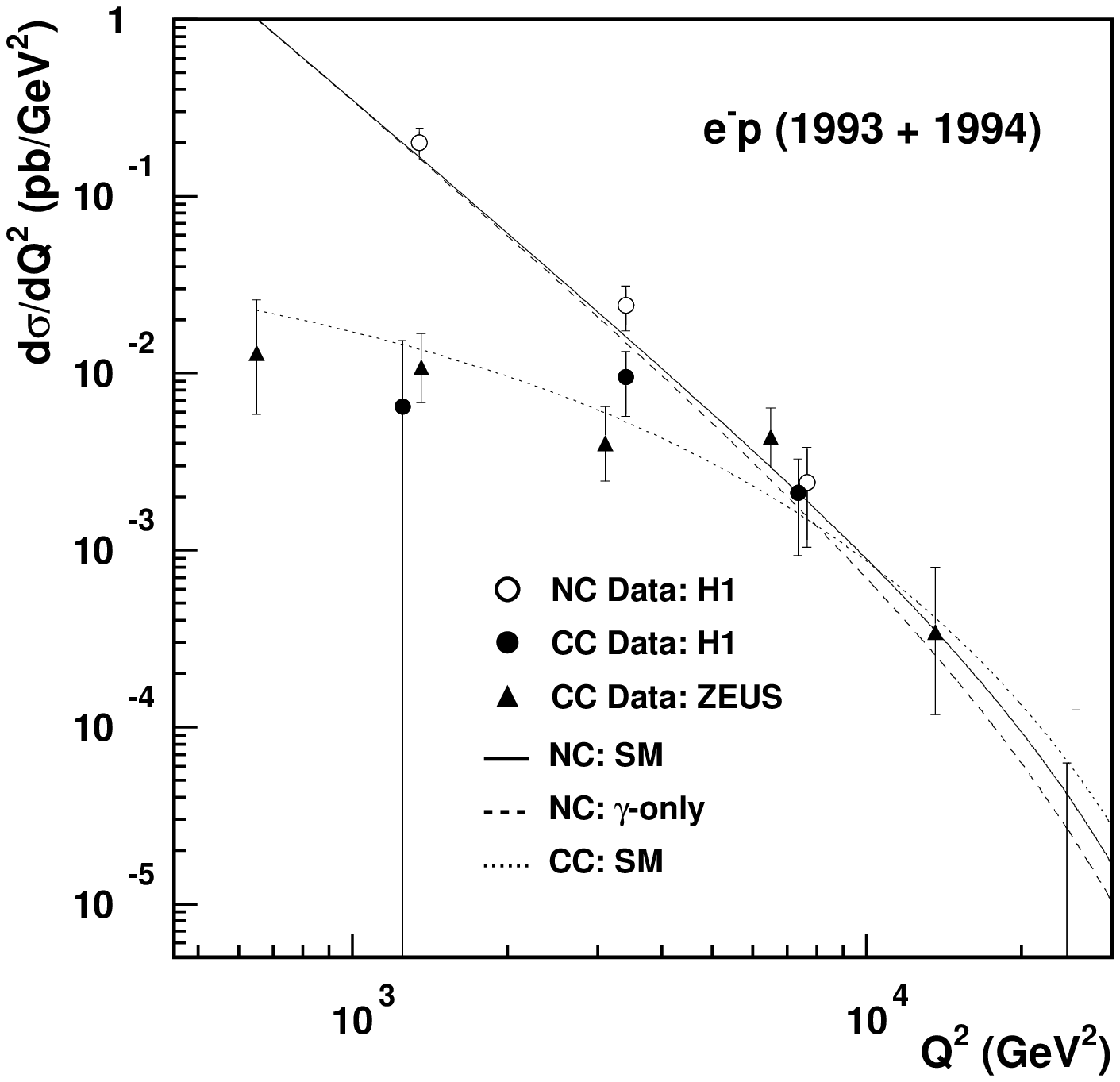,width=7.45cm,       
 bbllx=90pt,bblly=230pt,bburx=500pt,bbury=600pt}      
\epsfig{file=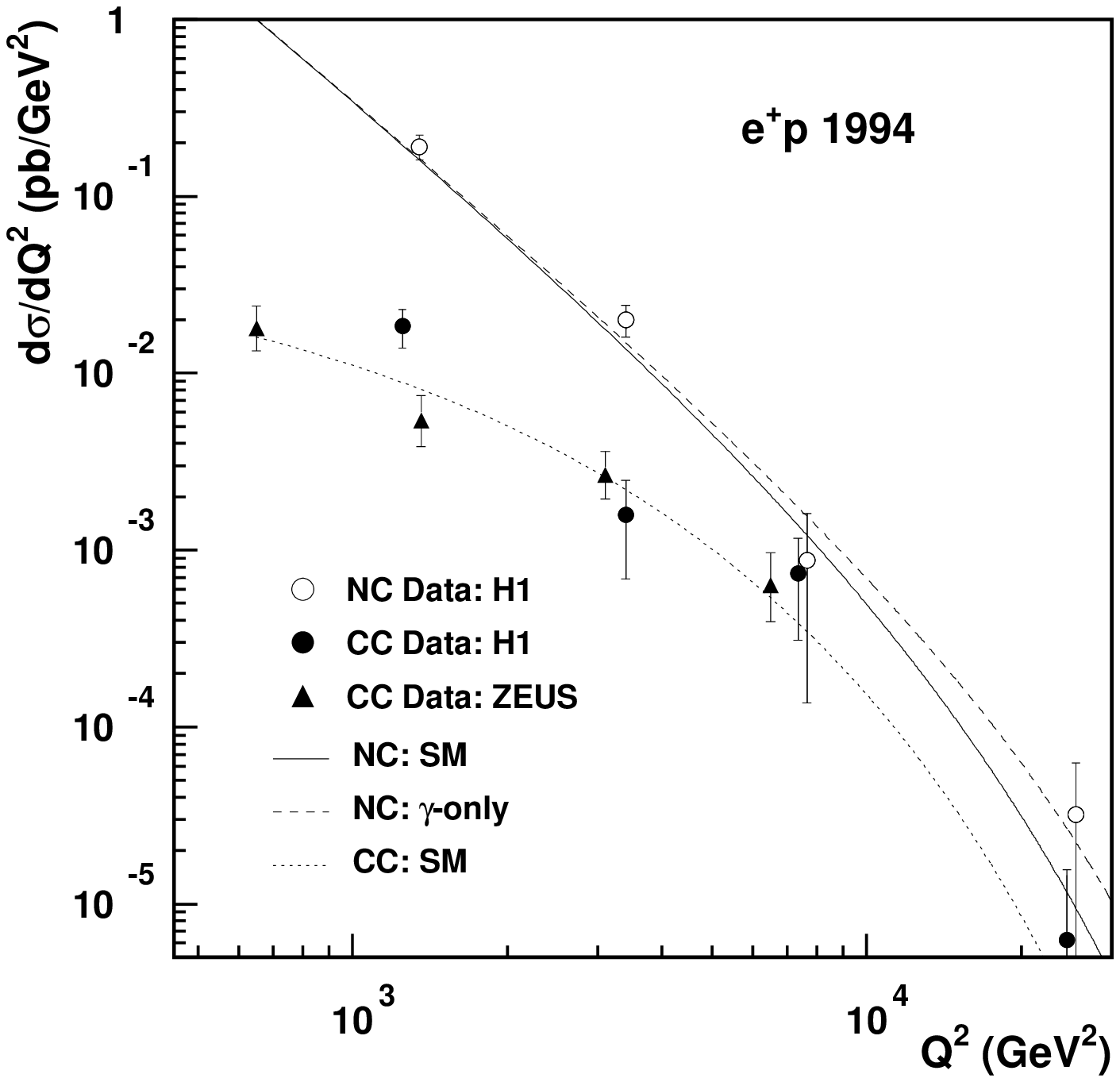,width=7.45cm,       
 bbllx=90pt,bblly=230pt,bburx=500pt,bbury=600pt}      
\end{center}      
\caption[]{\label{nccc}  
{\small \sl Measurement of the Born 
differential cross-section $d^2\sigma/dQ^2$ by
the H1 and ZEUS collaborations for NC and CC in $e^-p$ (a) and $e^+p$ (b)
collisions.
}}
\end{figure} 
At very high $Q^2$, i.e. when $Q^2 \simeq M_W^2$, the electroweak unification
is expected to become visible. At HERA, both
 charged (CC : $ep \rightarrow \nu X$)    and
neutral currents (NC : $ep \rightarrow e X$) 
are easily selected by requesting that the global
momentum transfer of the hadronic final state is
 large, typically  $\ge$ 20 GeV. The separation 
between CC and NC is based on the detection of the charged scattered lepton
present only in the NC events. In 1993 and 1994 a luminosity of
about 0.8 $pb^{-1}$ of $e^-p$ and 3 
$pb^{-1}$ of $e^+p$ collisions have been  recorded allowing new tests
of the Standard Model. The CC differential cross-sections
can be written in leading order as
\begin{eqnarray}
\frac{d^2\sigma^{e^-}_{CC}}{dQ^2} = \frac{G} {(1+Q^2/m_{W^-}^2)^2}
 [ u + c + (1-y)^2 (\bar{d} + \bar{s}+ \bar{b}) ] \\
\frac{d^2\sigma^{e^+}_{CC}}{dQ^2} = \frac{G} {(1+Q^2/m_{W^+}^2)^2} 
 [ d + s + b +(1-y)^2 (\bar{u}+ \bar{c}) ]
\end{eqnarray} 
 where G is a coupling constant, $m_{W^{\pm}}$ are the masses of the charged
weak bosons, and u,d,c,s,b, are the quark densities in the proton.
In fig.~\ref{nccc} the measurement of these differential cross-sections
are displayed and shown to be of a  similar strength as the NC one, when
at high $Q^2$. A shape and magnitude analysis of these distributions 
allows the  determination of the masses of the propagators involved. $M_{W^+}$
and  $M_{W^-}$ are found to be consistent, and a combined fit results in a 
mass $m_W$ = 84$^{+10}_{-7}$ GeV (H1,~\cite{H1CC94}) or 
79$^{+10}_{-7}$ GeV (ZEUS, preliminary), both  compatible with the precise  
measurement made at $p\bar{p}$ colliders where W's are directly produced.
The statistical error is still the dominant source of the total error given
on $M_W$ since only about 200 CC events have been recorded at HERA in 1993-94.
The effect of the $Z^0$ vector boson is not yet distinguishable from the 
single photon exchange cross-section, as shown in  fig.~\ref{nccc}.
With the expected higher luminosity, the ratio u/d and the
structure function $xF_3$ will be determined, while unexpected  
deviations from the Standard Model could reveal new insights in the deepest
structure of matter.

\section{Structure Function Measurement at HERA}    
In 1994 the H1 and ZEUS 
experiments have reduced the minimum $Q^2$ at which they   
could measure $F_2$ using several techniques:        
i) by diminishing the region around the backward 
beam pipe in which the electron could not be measured reliably in 1993,   
the maximum polar angle of the scattered electron (measured with respect to
the proton beam direction) was increased.       
The integrated luminosity of this large statistic sample is about  3~pb$^{-1}$.
ii) DIS events which underwent initial state     
photon radiation detected in an appropriate photon tagger were used       
to measure $F_2$ at lower $Q^2$ (so called ``radiative" sample)
since the incident electron energy  in the hard  
scattering is reduced.
iii) A luminosity of $\sim$ 
60~nb$^{-1}$ of data was collected for which the     
interaction point (vertex) was shifted by about +65~cm,      
in the forward direction,
resulting in an increase of the electron acceptance         
(so-called  ``shifted vertex" data sample).
The precision of these  luminosity measurements  are based on
the Bethe-Heitler reaction $e^-p \rightarrow e^-p \gamma$ and is   
1.5\% (3 to 4\% for the shifted vertex and radiative data).

The kinematic variables   
of the inclusive scattering process $ep \rightarrow eX$  
can be reconstructed in different ways using measured quantities from   
the hadronic final state and from the scattered electron.       
The choice of the reconstruction method for $Q^2$ and $y$       
determines the size of systematic errors, acceptance and radiative      
corrections. The measurements presented here have been obtained with the 
electron method (E),
with the $\Sigma$ method~\cite{sigma}, and with a combination of the 
double-angle (DA)~\cite{hoeger} and $\Sigma$ method.
The E method  has at large $y$ the best resolution
in $x$ and $Q^2$ but needs sizeable radiative corrections and cannot be used
at low $y$ due to the degradation
of the $y$ resolution as $ 1/y$. 
The $\Sigma$ method, which has small radiative corrections, relies
mostly on the hadronic measurement and can be used from very low
to large $y$ values.
H1 measures $F_2$ with the E and the $\Sigma$ method and after   
a complete consistency check, in particular at low $x$,
uses the E method for $y>0.15$      
and the $\Sigma$ method for $y<0.15$. ZEUS measures $F_2$ from the shifted vertex
and radiative data 
with the E method, otherwise with the $\Sigma$-DA combination.

The event selection is similar in the two experiments. Events are       
filtered on-line using calorimetric triggers which request an  
electromagnetic cluster of at least 5~GeV not vetoed by a trigger       
element signing a beam background event. Offline, further       
electron identification criteria are applied (track-cluster link,       
shower shape and radius) and a minimum energy of about 10 GeV is   
required. H1 requests a reconstructed vertex       
within 3$\sigma$ of the expected interaction position, while ZEUS       
requires, in order to reduce the photoproduction background and the      
size of the radiative corrections,
that  35~GeV $< \Sigma +E'_e(1-\cos\theta)<$ 65~GeV.        
The only significant background      
left after the selection comes from photoproduction in which    
a hadronic shower or a photon fakes an electron. In H1 for instance, 
it amounts to       
less than 3\% except in a few bins where it can reach values    
up to 15\%. It is subtracted statistically bin by bin and an error of   
30\% is assigned to it.

\begin{figure}[htb]   \unitlength 1mm      
\begin{center}    
\epsfig{file=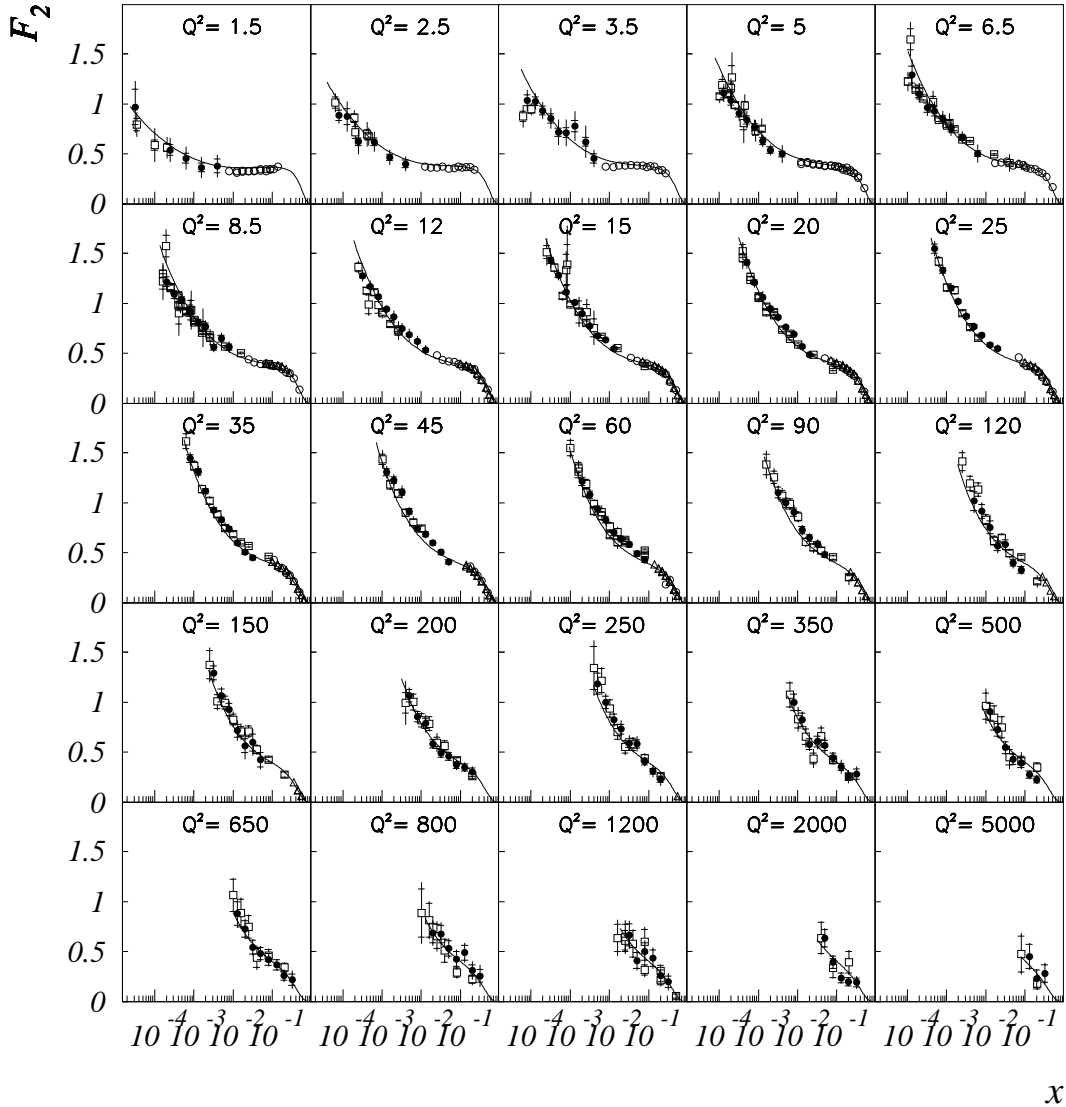,width=14.cm,       
 bbllx=35pt,bblly=170pt,bburx=550pt,bbury=690pt} 
\end{center}      
\caption[]{\label{f2q}  
{\small \sl $F_2(x,Q^2)$  measurement as a function
of $x$ by  H1 (black circles,\cite{H1F294}), 
 ZEUS (open squares, preliminary, except at low $Q^2$~\cite{ZEUSF294}), 
NMC (open circles, \cite{NMC}), BCDMS (open triangles, \cite {BCDMS}). The curve
represent the GRV model prediction.
}}
\end{figure}

The acceptance and the response of the detector have been studied and    
understood in great detail by the two experiments: 
more         
than two million  Monte Carlo DIS events, 
 corresponding to an 
integrated luminosity of  
approximately $20$~pb$^{-1}$,  were generated with    
DJANGO~\cite{DJANGO} using the
  GRV~\cite{GRV} and
 MRS~\cite{MRSA} parametrizations of the parton   
 distributions.
The Monte Carlo events, after a detailed detector simulation
were subjected to the same         
reconstruction and analysis chain as the real data.           

\begin{figure}[tbh]       
\begin{center}
\epsfig{file=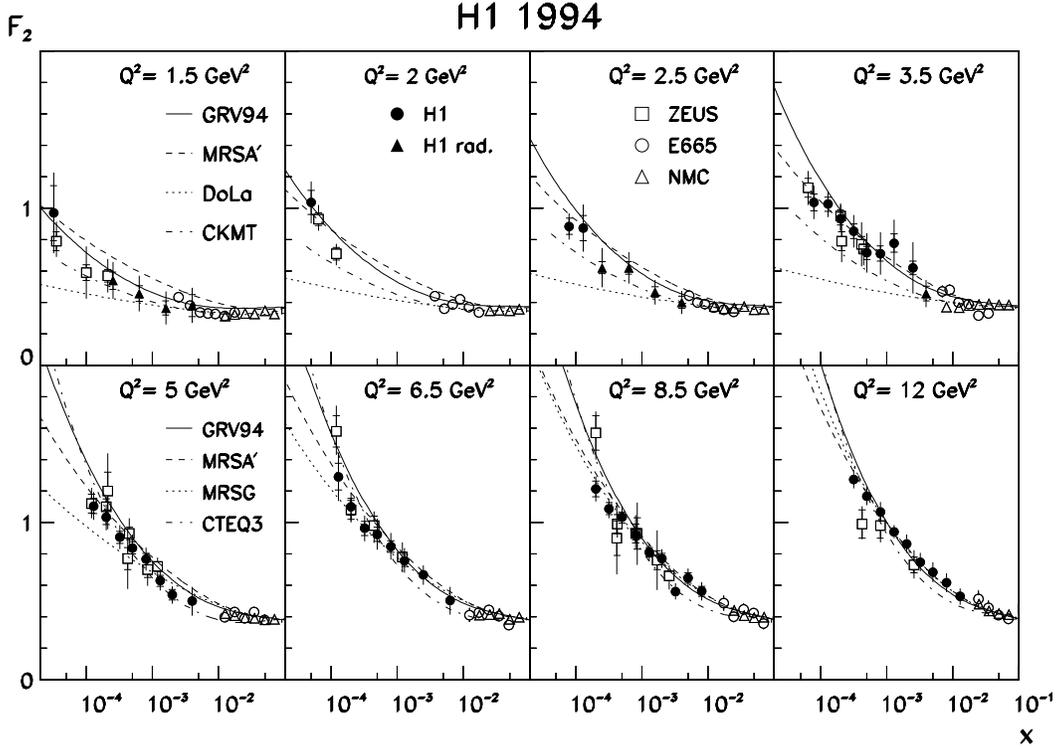,height=10.cm, 
 bbllx=40pt,bblly=240pt,bburx=540pt,bbury=600pt}      
\end{center}      
\caption[]{\label{lowq2}  
{\small \sl $F_2(x,Q^2)$  measurement
    in the low $Q^2$ region by H1 and ZEUS compared to         
    the results of the   
    E665 and NMC fixed target experiments. 
    The  $F_2$ parametrizations  confronted to the    
    data and discussed in the text are
 DOLA and CKMT at $Q^2 <$ 5 GeV$^2$, 
    GRV and MRSA'.}}        
\end{figure}

The  structure function $F_2(x,Q^2)$         
was derived after radiative corrections      
from the differential cross-section $d^2 \sigma/dx dQ^2$.
($R \equiv F_2/2xF_1 - 1 $ was taken as prescribed 
by QCD).      
With  the different data sets available, 
detailed  cross  checks could be made in the kinematic regions of overlap.   
The results were found to be in  very good  agreement   
with each other for all kinematic reconstruction methods used, and
the effect of systematic errors could be monitored. 
For the E method the main source of error are the energy calibration 
(known at the  1\% level), the knowledge of the electron   
identification efficiency,
the error on the polar 
angle of the scattered electron (1 mrad), 
and the radiative corrections at         
low $x$.  
The DA method becomes strongly sensitive to the precision of the 
 of the electron and  hadronic angle
when they tend to their boundary values (0 or $\pi$). 
For the $\Sigma$ method, the knowledge of the absolute energy scale     
for the hadrons, the fraction of hadrons which stay undetected,  
in particular at low $x$, due to calorimetric thresholds        
and to a lesser extent the electron energy calibration are the  
dominating contributions.
The total $F_2$ 
errors on the 1994 data ranges between 5 and 10\% in the  10-100 GeV$^2$ 
range and between 10 and 20\% below 10 GeV$^2$.      
The final results from the 1994 data
 of H1 and ZEUS are shown  in fig.~\ref{f2q}.
Compared to the 1993 data analyses
the $F_2$ measurement has been extended 
to lower $x$
(from $1.8 \cdot 10^{-4}$ to   
$3.5 \cdot 10^{-5}$)   and in      
$Q^2$ (from 4.5-1200~GeV$^2$ to  1.5-5000~GeV$^2$).        
Both experiments are in good agreement and show that the $F_2$ rise at low
$x$ persists, albeit less strongly, down to the lowest measured $Q^2$=1.5 
GeV$^2$. 
A smooth transition between the HERA data
and the results of the fixed target experiments NMC~\cite{NMC} and 
BDMCS~\cite{BCDMS} 
is observed, and for the first time also at low $Q^2 \le$
5 GeV$^2$ between E665~\cite{E665} and the low $y$ results of H1
 (fig~\ref{lowq2}),
allowing  these consistent results to be confronted with 
theoretical expectations.

\section{Structure Functions at Low $Q^2$}
In fig.~\ref{lowq2} 
we focus on the low $Q^2$ measurements, i.e. 
in the new kinematic domain reached
using the radiative and the shifted 
vertex data. 
Also  shown  are the extrapolations of the $F_2$ 
parametrizations based on theoretical models 
adjusted to the  previous data.  They can be divided in two categories:  one, 
motivated by Regge theory, assumes pomeron exchange as a dynamical
basis and successfully describes the behaviour of the total cross-sections  
of photoproduction and hadron-hadron collisions;  the other is based on
perturbative QCD and is known to  describe well the evolution of the
DIS cross-sections, but
is expected to fail at some low $Q^2$.

\begin{figure}[htbp]    \unitlength 1mm
\begin{center}
\begin{picture}(150,75)
\put(-3,0){
\begin{picture}(0,0) \put(15,70){({\large \bf a})} \end{picture}
\begin{picture}(0,0) \put(90,70){({\large \bf b})} \end{picture}
\epsfig{file=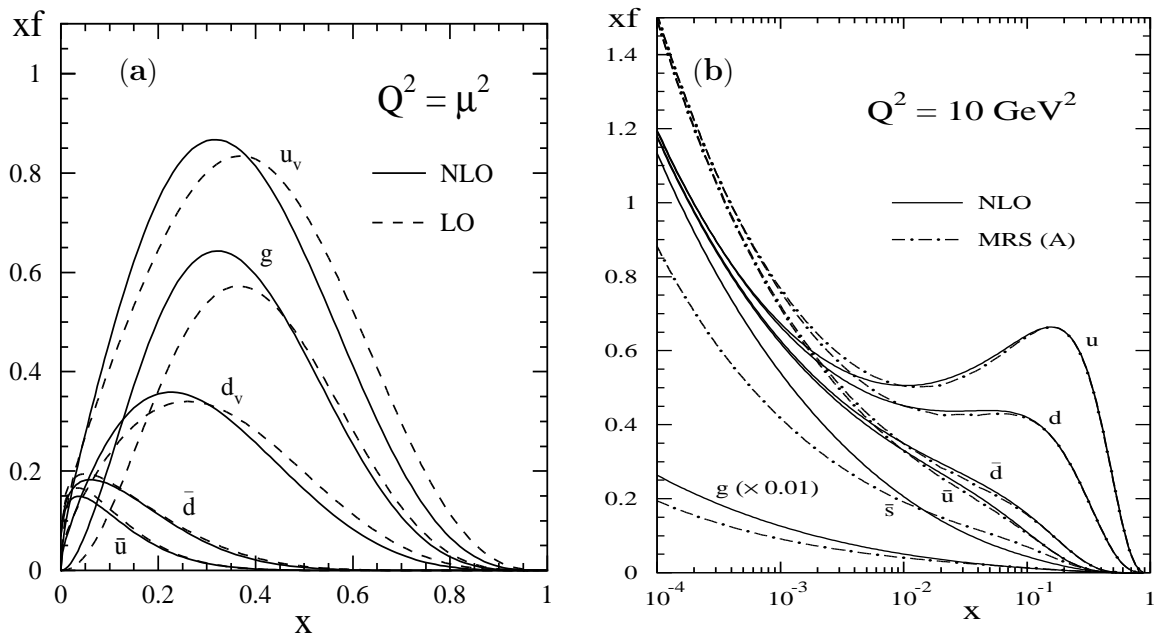,height=7.2cm,
width=7.5cm,bbllx=150pt,bblly=260pt,bburx=480pt,bbury=570pt}  
\epsfig{file=fig4b.ps,height=7.5cm,
width=7.5cm,bbllx=55pt,bblly=497pt,bburx=290pt,bbury=810pt}   
}
\end{picture}
\end{center}   
\caption[]{\label{glugrv}      
{\small \sl 
a) Parton densities (valence quarks (u$_v$,d$_v$),
gluon (g) and sea quarks) of the GRV model at the initial energy scale
$\mu^2=0.34$ GeV$^2$.
b) at a scale of 10 GeV$^2$, and compared to  MRSA.}}
\end{figure} 

\begin{figure}[htbp]       
\begin{center}    
\begin{picture}(150,75)
\put(-3,0){
\begin{picture}(0,0) \put(10,12){({\large \bf a})} \end{picture}
\begin{picture}(0,0) \put(60, 5){({\large \bf b})} \end{picture}
\begin{picture}(0,0) \put(112, 5){({\large \bf c})} \end{picture}
\epsfig{file=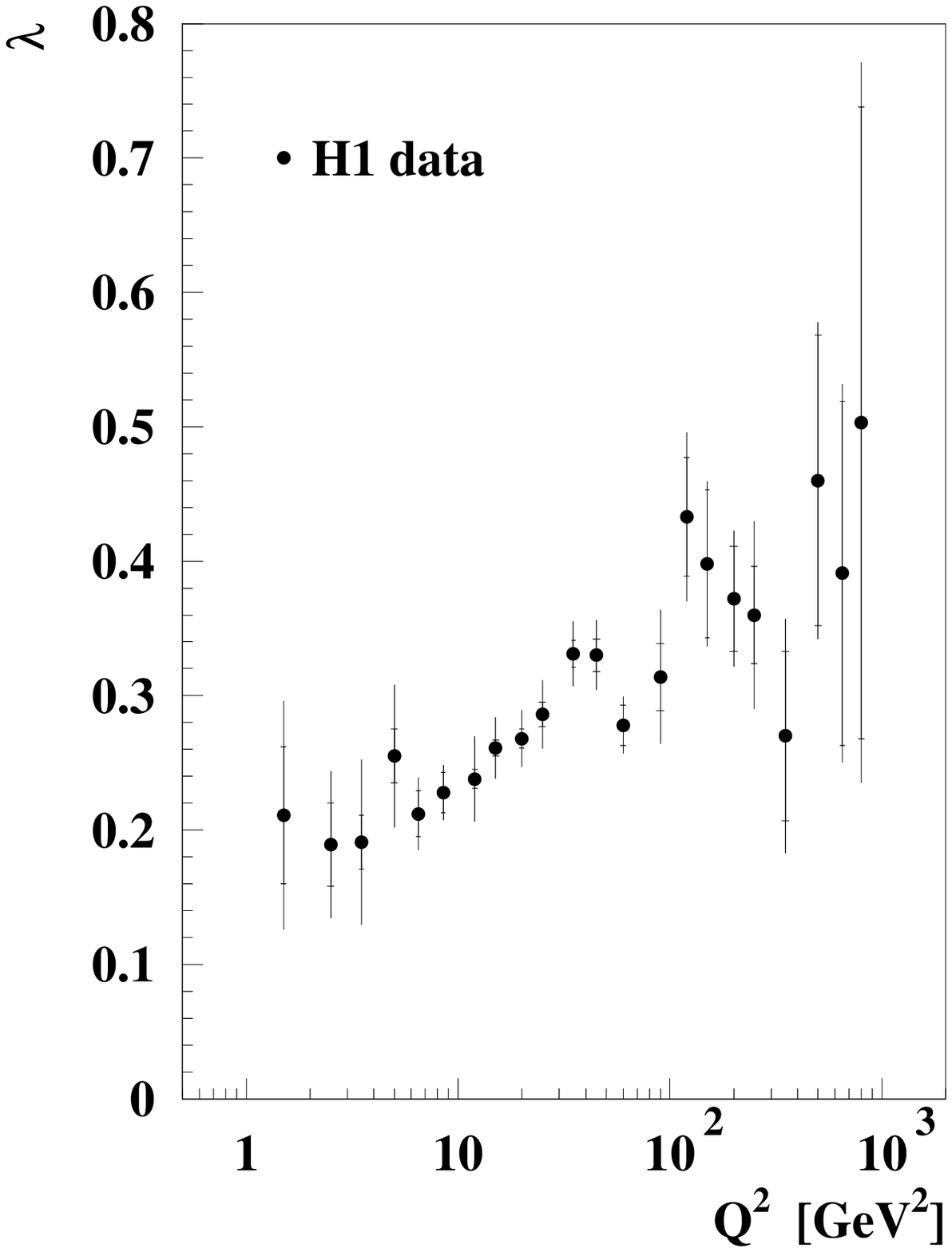,
width=5.cm,bbllx=120pt,bblly=140pt,bburx=490pt,bbury=660pt}
\epsfig{file=fig5b.ps,width=9.8cm,       
 bbllx=100pt,bblly=460pt,bburx=440pt,bbury=710pt}
} 
\end{picture}     
\end{center}      
\caption[]{\label{e665}
{\small \sl a) 
Variation of $\lambda$ obtained from fits of the form $F_2 \sim x^{-\lambda}$
to the H1 data, at fixed $Q^2$ and $x<$0.1. 
b), c)
 $F_2(x,Q^2)$ measurement
    in the low $Q^2$ region by the  E665  experiment. 
    Different  $F_2$ parametrizations are confronted to the
    data:  DOLA , BK and  GRV (see text).}}        
\end{figure} 

The Regge models were expected to work at least at low $Q^2$, but the
DOLA parametrization which uses a ``soft'' pomeron 
(intercept$\simeq$ 1.08)~\cite{DOLA} largely underestimates $F_2$ at 
low $x$ even at 1.5 GeV$^2$.      
The CKMT model~\cite{CKMT}, which    
assumes that in the present  $Q^2$ range
the ``bare" pomeron becomes visible and
has a higher     
trajectory intercept ($\simeq 1.24$),
predicts a weaker rise at low $x$  than observed, except maybe at
 1.5 GeV$^2$.  
These comparisons underline the difference between the behaviour of the total
cross-section of real and virtual photons, since in the HERA kinematic
domain 
 $\sigma_{tot}^{\gamma^* p}$ can be expressed as
\begin{equation}  
\sigma_{tot}^{\gamma^* p}(x,Q^2) \simeq (4~\pi^2 \alpha /Q^2 )\cdot
 F_2(x,Q^2). 
\end{equation}

The parametrizations based on the DGLAP QCD evolution equations describe the
data remarkably well, as expected above 5-10 GeV$^2$, but also surprizingly
at values around 1 or 2 GeV$^2$ where non-perturbative effects were 
believed to distort the DGLAP picture. 
The MRSA' parametrizations of the 
parton densities are defined at $Q^2_0=$ 4 GeV$^2$, then 
evolved in $Q^2$ and fitted to previous experimental data, including the 1993
HERA data. The agreement observed above 10 GeV$^2$ confirms that the 1993 and
1994 HERA results are perfectly compatible. Between 1.5 and 10 GeV$^2$ the good
description  tells us that within the present precision perturbative QCD
can be applied in this range.
More striking is the confirmation of the pre-HERA prediction of the $F_2$
rise at low $x$ by the
GRV model~\cite{GRV}
 which conjectured that at a very low energy scale ($\mu^2$=0.34 GeV$^2$)
the proton is formed by valence-like partons as shown in 
fig.~\ref{glugrv}a
and that the DGLAP equations can be applied to generate ``radiatively'' the
rise of the gluon and sea-quark density at low $x$, when evolving towards 
higher $Q^2$ (fig.~\ref{glugrv}b). 
The H1 and ZEUS results are well  described by the GRV model 
as can be seen in fig.~\ref{f2q} and \ref{lowq2}.
This success supports the idea that the rise at low $x$ is a direct 
consequence of the DGLAP equations, and that non-perturbative effects
are relatively weak at low $x$ and low $Q^2$.

The evolution with $Q^2$ of the strength of the rise can be quantified
by fitting an $x^{-\lambda}$ function 
(or equivalently a form $W^{2\lambda}$, $W$ being
the invariant mass of the $\gamma^{\star}-p$ system,
as shown in~\cite{Levy}) at fixed $Q^2$
to $F_2(x), \ x < 0.1$. The values of $\lambda$ obtained by the fit in 
each $Q^2$ bin are displayed in fig.~\ref{e665}a and clearly confirm 
the prediction of many years'standing made 
for asymptotic free field theories like
QCD~\cite{ALVARO} of a rise of $F_2$ at low $x$, and that
the size of this rise increases with $Q^2$. 
With the present data, it is however not possible
to know  precisely this size below 5 GeV$^2$, thereby postponing
a definite test of perturbative QCD in this region. 

Recent results from the fixed target E665 experiment provide additional 
information, in particular below 5 GeV$^2$,
while at $Q^2 \ge$ 5 GeV$^2$ the E665 and NMC measurements are in good
agreement (fig.~\ref{lowq2}). The $x$ range however does not extend
to the HERA values since its limit varies between 10$^{-2}$ and 10$^{-3}$.
In fig.~\ref{e665}b,c the E665 data are shown to be described in this medium
$x$ range and for $Q^2 >$ 0.3  GeV$^2$ by the DOLA and BK~\cite
{BK}  
parametrizations, the latter being based on the concept of Generalized Vector
Meson Dominance (GVMD) at low $Q^2$ with a smooth transition to perturbative
QCD at higher $Q^2$. At values below 0.7 GeV$^2$ the GRV description starts
to break down,
but the GRV approach is  still  valid 
below 1 GeV$^2$. The new HERA data taken in 1995 with upgraded detectors
will further constrain these models at low $x$ for
$Q^2 <$ 1 GeV$^2$ thereby checking if perturbative QCD  can indeed explain
the dynamics at these very low $Q^2$.

\begin{figure}[htbp]    \unitlength 1mm
\begin{center}
\epsfig{file=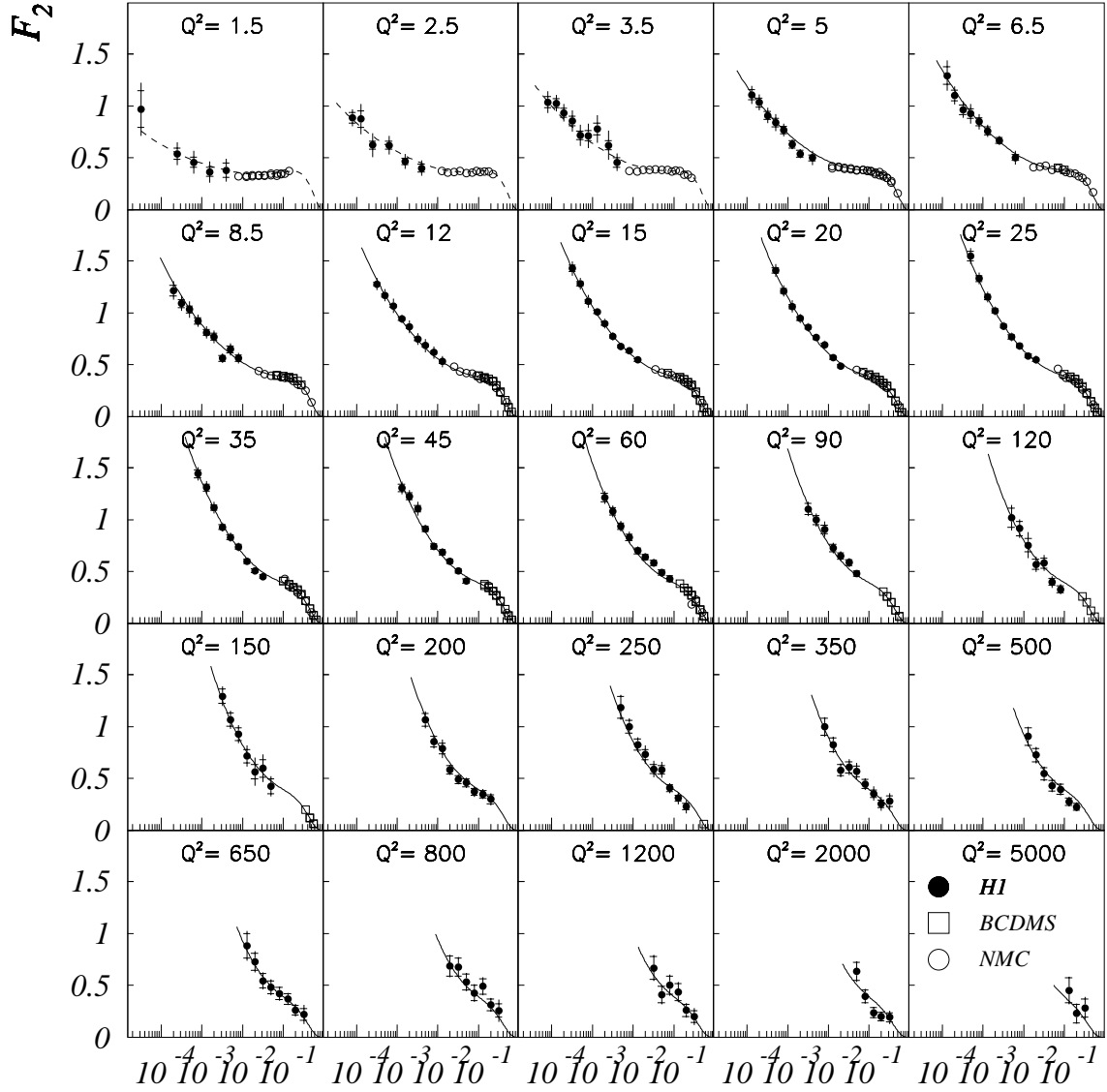,width=15.cm,    
 bbllx=35pt,bblly=170pt,bburx=550pt,bbury=690pt} 
\end{center}      
\caption[]{\label{f2qfit}    
{\small \sl $F_2(x,Q^2)$ measurement as a function of $Q^2$.
The curve represent a NLO QCD fit to the H1 (black circles), BCDMS (open
   squares) and NMC (open circles) data at Q$^2 > $5 GeV$^2$.
}}       
\end{figure}

\section{Structure Functions and Perturbative QCD}     
\begin{figure}[htb]    \unitlength 1mm
\begin{center}
\begin{picture}(150,75)
\put(-5,0){
\begin{picture}(0,0) \put(10,75){({\large \bf a})} \end{picture}
\begin{picture}(0,0) \put(85,75){({\large \bf b})} \end{picture}
\epsfig{file=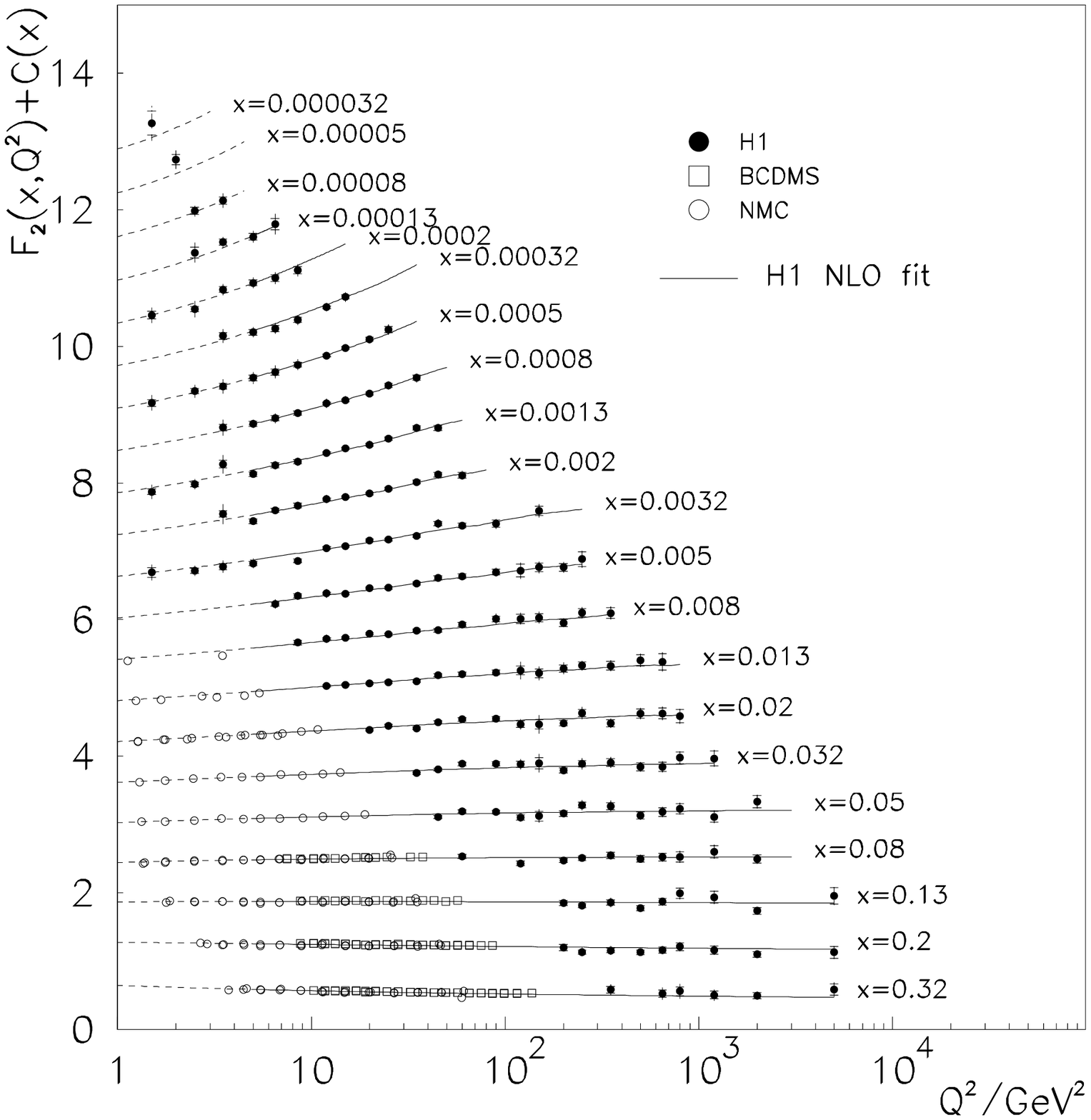,
width=7.5cm,bbllx=25pt,bblly=160pt,bburx=570pt,bbury=685pt}
\epsfig{file=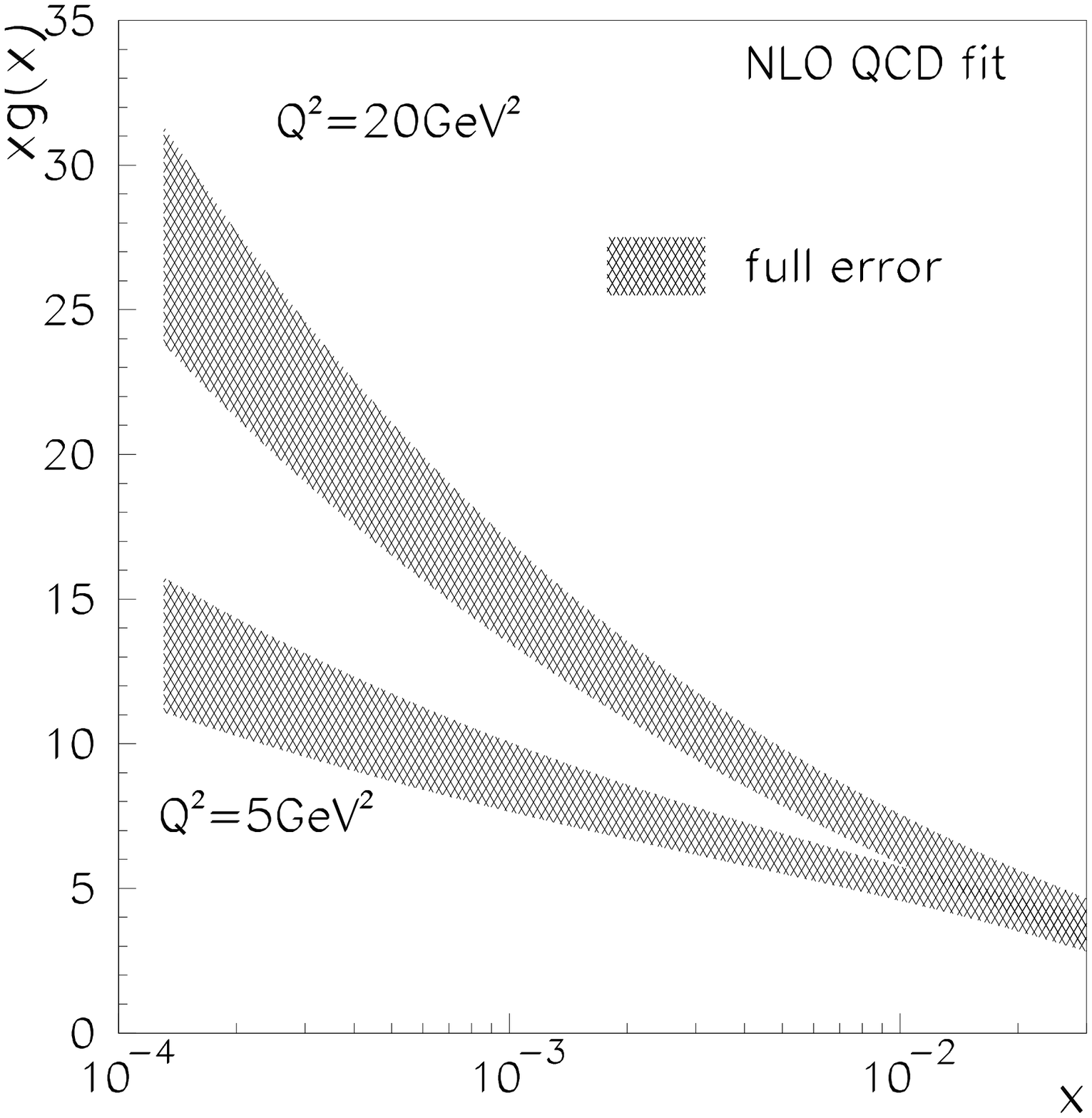,
width=7.5cm,bbllx=25pt,bblly=160pt,bburx=570pt,bbury=685pt}   
}
\end{picture}
\end{center}   
\caption[]{\label{f2xfit}      
{\small \sl 
a) H1, BDCMDS and NMC  measurement 
    of the proton structure function $F_2(x,Q^2)$   
    as function of  $Q^2$. The curve represent the NLO QCD fit to the
H1, BCDMS and NMC data described in the text.
b) Gluon density at
5 and 20 GeV$^2$ determined by a NLO fit
to the H1,NMC and BCDMS data. The error bands represent the full error
except for the uncertainty on $\Lambda$.}}
\end{figure}

To make full use of the new precision achieved with the 1994 data, the H1
collaboration has performed 
a Next-to-Leading Order (NLO) QCD fit on the H1, BCDMS and NMC
data with the conditions
$Q^2 >$ 5 GeV$^2$, and $x <$ 0.5 if $Q^2 <$ 15 GeV$^2$ to avoid 
higher-twist effects.
The H1 measurements which extend  to  5000 GeV$^2$ 
were fitted successfully (fig.~\ref{f2qfit}) and
 constrain the gluon density at low $x$. The parton densities were
parametrized at $Q^2_0$=5 GeV$^2$, in particular the gluon was expressed
with 3 parameters as $xg(x)= A_gx^{B_g}(1-x)^{C_g}$.
The quark and antiquark components of the sea were assumed to be equal,
and $\bar{u}$ set equal to $\bar{d}$. As determined in ~\cite{ccfr},
the strange quark density was taken to be $\bar{s}=(\bar{u}+\bar{d})/4$.
Further constraints due to  quark counting rules and 
momentum sum rules
were included. For $\Lambda$ the value of 263 MeV was taken~\cite{qcdbcd}.
A detailed treatment of the $F_2$ error propagation on the 
gluon density has been made, resulting in the error bands of fig.~\ref{f2xfit}b
which represent $xg(x)$ at 5 and 20 GeV$^2$. 
A variation of 
$\Lambda$ by  65 MeV 
gives a  change of 9\%  on the gluon density 
at 20 GeV$^2$ which has not been added to the error bands.
The accuracy  of this determination of $xg$  
is better by  about a factor of two  than  the H1 result based on
the 1993 data~\cite{qcdfit}.
A  rise of the gluon density towards low $x$ is observed
which  is related  to the behaviour of $F_2 \propto x^{-\lambda}$.
Accordingly, the rise of $xg$ towards low $x$ increases with increasing $Q^2$.
Finally we can observe in fig.~\ref{f2qfit} 
that the data at 
$Q^2 < 5 $~GeV$^2$, which were excluded from the fit, are still
well reproduced by the fit evolved backwards in $Q^2$.
More data at low $x$ and $Q^2 < 1
 $~GeV$^2$ are nevertheless needed to be able
to test the hypothesis of a gluon density which would take the valence-like
shape displayed in fig.~\ref{glugrv}a when $Q^2 \rightarrow$ 0.3 GeV$^2$,
 and more generally, to better understand the dynamics at low $Q^2$ and 
low $x$ were
parton densities are high.
The HERA experiments, which have last year upgraded
their backward detectors,  will be able to reach such low 
$Q^2$ with the data taken in 1995 and 1996.

In conclusion, the HERA 1994 data with their improved precision have            
provided new tests which have been passed successfully by                       
perturbative QCD on 3 orders of magnitude in $Q^2$, between 5 and                 
5000    GeV$ 2$. Another test not described in this                             
report is the observation by the H1 collaboration~\cite{H1F294}               
of double asymptotic scaling~\cite{ball}  as predicted by QCD, 
 for $Q^2$    values above 5 GeV$^2$.                                                          
At the present level of precision the DGLAP evolution equations                 
are sufficient                                                                  
to account for the observed rise of $F_2$ at fixed $Q^2$, although it            
is not yet possible to distinguish between the different solutions                
of the ``input'' parton distribution problem which allow for a good            
description of the data.                                                        
From 1.5 to 5 GeV$^2$ all analyses/interpretation of the $F_2$ behaviour          
hint that perturbative QCD is also applicable at these low $Q^2$.                
However the lower statistical                                                   
precision of these measurements obtained with                                   
dedicated data samples prevent a definite conclusion at the moment.             
The forthcoming 1995 results should have the precision and the                  
extension at even lower $Q^2$ sufficient to constrain the limit                  
of validity of perturbative QCD whose domain has been already observed          
to be wider than generally expected.                                            

\newpage
\noindent
{\Large{{\bf Acknowledgements}}}    

\normalsize       
\vspace*{0.1cm}
\noindent         
I would like to thank the organizers, particularly Ben Gibson, 
to have realized  such an interesting conference in the nice town of
Santa F\'e.
I would also like to thank my close collaborators,
U. Bassler, B. Gonzalez-Pineiro and F. Zomer, J. Dainton
for a careful reading of the manuscript,
all the colleagues of the
H1 structure 
function group in particular A. DeRoeck, J. Feltesse and M. Klein,  
and the E665, H1 and ZEUS collaboration
with whom we obtained the recent results
described above.
     
\end{document}